\documentclass[conference]{IEEEtran}
\IEEEoverridecommandlockouts
\usepackage{cite}
\usepackage{amsmath,amssymb,amsfonts}
\usepackage[a4paper, total={170mm,257mm},
left=30mm,
top=30mm,
right=30mm,
bottom=32mm]{geometry}
\usepackage{algorithmic}
\usepackage{graphicx}
\usepackage{textcomp}
\usepackage{xcolor}
\usepackage{booktabs,tabularx}
\usepackage{hyperref}
\usepackage{url}
\usepackage{multirow, multicol}
\def\BibTeX{{\rm B\kern-.05em{\sc i\kern-.025em b}\kern-.08em
    T\kern-.1667em\lower.7ex\hbox{E}\kern-.125emX}}
\begin{document}

\title{The Turing Synthetic Radar Dataset: A dataset for pulse deinterleaving.\\

\thanks{
This work was supported by the Turing's Defence and Security programme through a partnership with the UK government in accordance with the framework agreement between HMG and The Alan Turing Institute. We acknowledge the contributions of the broader electronic warfare research community and the participants who will contribute to making this challenge successful.

Particularly we would like to acknowledge Leonardo for their contributions to designing the dataset and Marco Fontana for discovering various features of the dataset.}
}

\author{\IEEEauthorblockN{1\textsuperscript{st} Edward Gunn}
\IEEEauthorblockA{\textit{The Alan Turing Institute} \\
London, The United Kingdom \\
egunn@turing.ac.uk}
\and
\IEEEauthorblockN{2\textsuperscript{nd} Adam Hosford}
\IEEEauthorblockA{\textit{Dstl} \\
Porton Down, The United Kingdom \\
ahosford@mail.dstl.gov.uk}
\and
\IEEEauthorblockN{3\textsuperscript{rd} Robert Jones}
\IEEEauthorblockA{\textit{The Alan Turing Institute} \\
London, The United Kingdom \\
rjones@turing.ac.uk}
\and
\IEEEauthorblockN{4\textsuperscript{th} Leo Zeitler}
\IEEEauthorblockA{\textit{The Alan Turing Institute} \\
London, The United Kingdom \\
lzeitler@turing.ac.uk}
\and
\IEEEauthorblockN{5\textsuperscript{th} Ian Groves}
\IEEEauthorblockA{\textit{The Alan Turing Institute} \\
London, The United Kingdom \\
igroves@turing.ac.uk}
\and
\IEEEauthorblockN{6\textsuperscript{th} Victoria Nockles}
\IEEEauthorblockA{\textit{The Alan Turing Institute} \\
London, The United Kingdom \\
vnockles@turing.ac.uk}
}

\maketitle

\begin{abstract}
We present the \textit{Turing Synthetic Radar Dataset}, a comprehensive dataset to serve both as a benchmark for radar pulse deinterleaving research and as an enabler of new research methods. The dataset addresses the critical problem of separating interleaved radar pulses from multiple unknown emitters for electronic warfare applications and signal intelligence. Our dataset contains a total of 6000 pulse trains over two receiver configurations, totalling over 4 billion pulses, featuring realistic scenarios with up to 90 emitters and significant parameter space overlap. To encourage dataset adoption and establish standardised evaluation procedures, we have launched an accompanying \textit{Turing Deinterleaving Challenge}, for which models need to associate pulses in interleaved pulse trains to the correct emitter by clustering and maximising metrics such as V-measure. The \textit{Turing Synthetic Radar Dataset} is one of the first publicly available, comprehensively simulated pulse train datasets aimed to facilitate sophisticated model development in the electronic warfare community. 
\end{abstract}


\section{Introduction}
\begin{figure*}[!ht]
    \centering
    \includegraphics[width=0.75\linewidth]{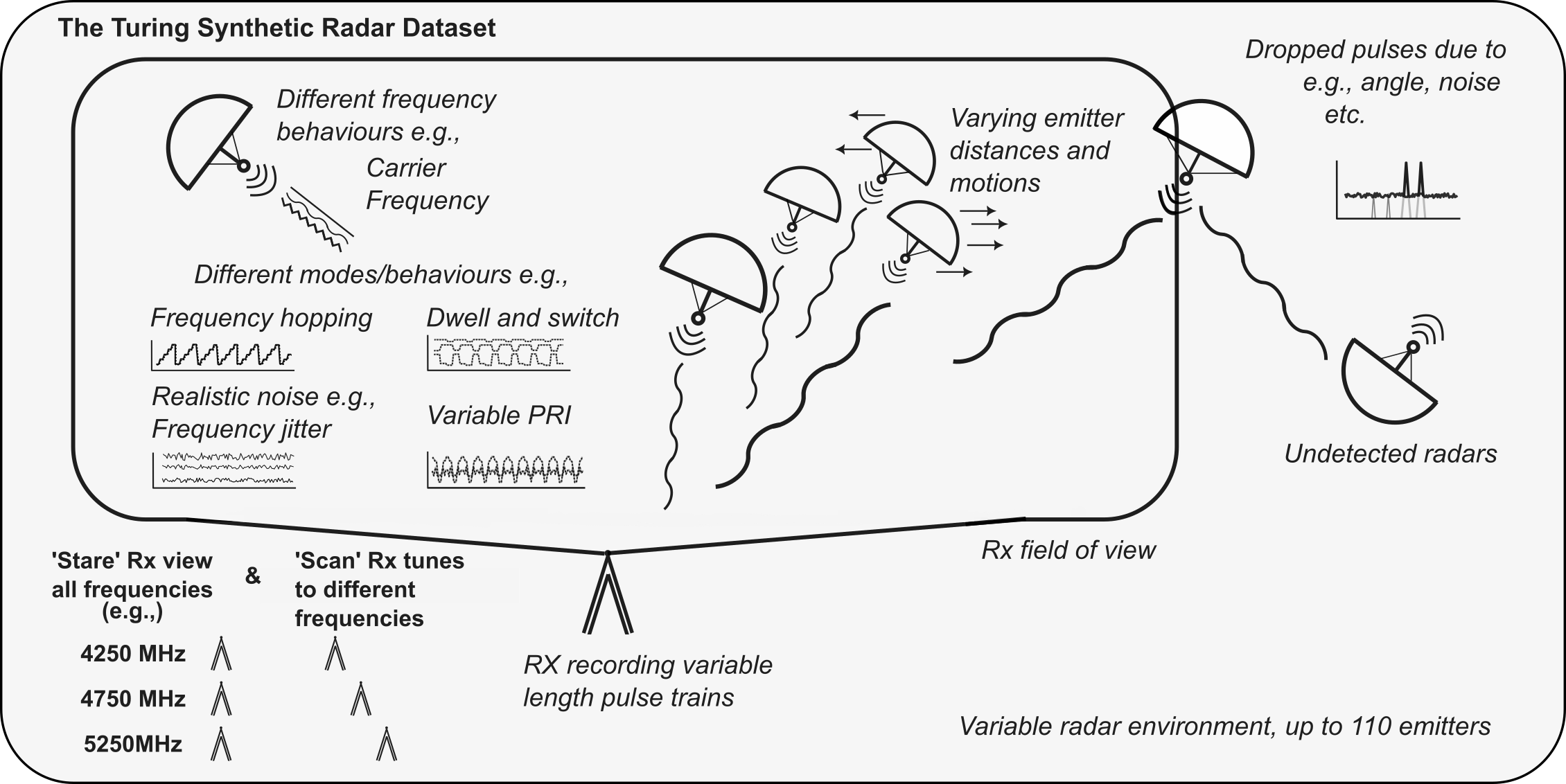}
    \caption[The TSRD includes realistic transmitter-receiver behaviours.]{\textbf{The TSRD includes realistic transmitter-receiver behaviours.} For each simulated pulse train, a static receiver detects pulses from multiple emitters at varying distances on a two-dimensional plane, simulating realistic signal propagation effects, such as path loss and detected angle of arrival. Pulses sent from too far or at the wrong angle are not detected. Emitters operate on different modes, which includes the pulse frequency intervals, frequency modulations, and other advanced techniques.}
    \label{fig:summary-schematic}
\end{figure*}
Radar pulse deinterleaving is a fundamental challenge in radar signal analysis and electronic warfare (EW) systems, where the task is to separate interleaved pulse trains received from multiple unknown radar emitters \cite{qu_intelligent_2025}. This problem is critical for signal analysis, enabling downstream tasks such as specific emitter identification, radar mode classification, and threat assessment. The modern radar environment is characterised by congested spectrum and increasingly agile radar systems, which pose significant challenges that exceed the capabilities of traditional approaches \cite{gunn_radar_2025}.

The deinterleaving problem requires partitioning a sequence of radar pulses by their originating emitters, where the number of emitters is unknown and varies significantly between pulse trains. Each pulse is characterised by a pulse descriptor word (PDW) containing features such as time of arrival, frequency, pulse width, angle of arrival, and amplitude. The challenge lies in identifying patterns and correlations within these features that reliably distinguish between different emitters while handling realistic complications such as missing pulses, parameter variations, and adversarial behaviour.

Current research in radar pulse deinterleaving faces several limitations. Many existing approaches assume a fixed number of emitters, reducing the problem to classification rather than the more realistic clustering scenario. Others focus primarily on pulse repetition interval (PRI) analysis while ignoring other valuable PDW features \cite{nuhoglu_radar_2023, xie_novel_2023}. The lack of standardised datasets and evaluation metrics makes it difficult to compare different approaches and track progress in the field \cite{gunn_radar_2025, qu_intelligent_2025}.

\par The shortage of publicly available datasets for EW-related model development has been highlighted previously \cite{huang2024multi, reddy2025state}. Although there are several radio datasets \cite{o2018over, jagannath2021multi, huang2023multi, clerico2023lstm}, they are largely directed at radio communication applications, and they are insufficient for congested radar environments. Moreover, the data are deinterleaved and sometimes incompletely annotated IQ streams of short duration. Addressing this, \cite{huang2024multi} published an IQ dataset of interleaved sequences. However, IQ streams are still limited to relatively short emissions of 10.24 $ms$. Instead of focusing on IQ data, \cite{sun2025semi} constructed sequences of PDWs, which allows simulating longer time scales than IQ data. Unfortunately, their data is only available from the author \textit{upon reasonable request}. It should be noted that all of these datasets were published together with a model, rather than focusing on developing a model-independent dataset for complex superimposed radar pulse trains.

To address these issues, we introduce \textit{The Turing Synthetic Radar Dataset} (TSRD), the first publicly available and model-agnostic dataset of its kind for the radar deinterleaving community. Rather than focusing on IQ data streams, the TSRD is composed of sequences of PDWs which allows simulating much larger and more complex pulse trains. Our aim is to provide a benchmark dataset which contains i) realistic synthetic data at scale which takes physical transmission properties into account; ii) standardised evaluation metrics; and iii) enables multiple avenues of applied research into state-of-the-art deinterleaving. Whilst our dataset is purely synthetic, we have engineered a large diversity of real-world scenarios, guiding our simulations to produce real-world complexity, while providing the necessary ground truth labels necessary for supervised learning and objective evaluation.

To summarise, we make the following contributions:
\begin{itemize}
    \item Implementing a pulse train generation pipeline with realistic variation in the degree of complexity.
    \item Publishing a first-of-its-kind synthetic radar dataset for model development of interleaved pulse sequences
    \item Proposing the \textit{Turing Deinterleaving Challenge} as new benchmark for both new and existing computational models that ingests interleaved PDW sequences from different emitters and deinterleaves them.
\end{itemize}
In the following sections, we describe our approach to generating realistic synthetic radar data, detail the structure and characteristics of the dataset (Section \ref{sec:headings}), and outline the evaluation framework for the accompanying challenge (Section \ref{sec:outlook}).
The dataset and documentation are publicly available at \url{https://huggingface.co/datasets/alan-turing-institute/turing-deinterleaving-challenge}
\section{Generating a Realistic Radar Pulse Train Dataset}
\label{sec:headings}
\begin{figure}[!ht]
    \centering
    \includegraphics[width=1.0\linewidth]{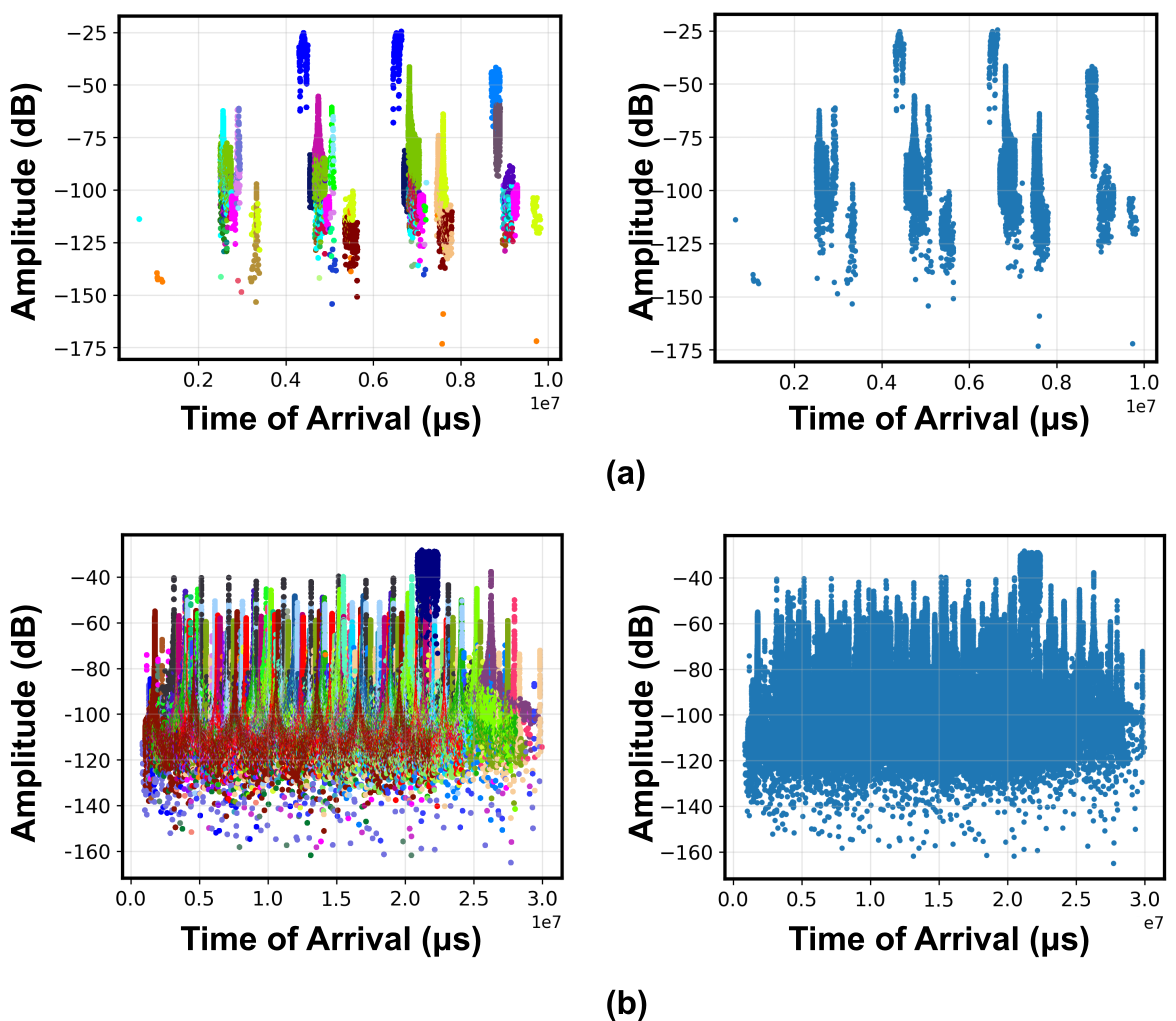}
    \caption[Emitted pulses substantially overlap in the parameter space, rendering straightforward deinterleaving challenging.]{\textbf{Emitted pulses substantially overlap in the parameter space, rendering straightforward deinterleaving challenging.} (A) and (B) exemplify two received pulse trains over ToA and amplitude in \textit{scan} and \textit{stare} mode, respectively, demonstrating that emitter signals are substantially superimposed. Simple deinterleaving is challenging, requiring sophisticated model development that makes use of clean data with ground truth labels (represented by the colours in the left panels).}
    \label{fig:trains-with-and-without-labels}
\end{figure}
\begin{figure*}[ht!]
    \centering
    \includegraphics[width=0.7\linewidth,trim={1.5cm .5cm 1.5cm 0},clip]{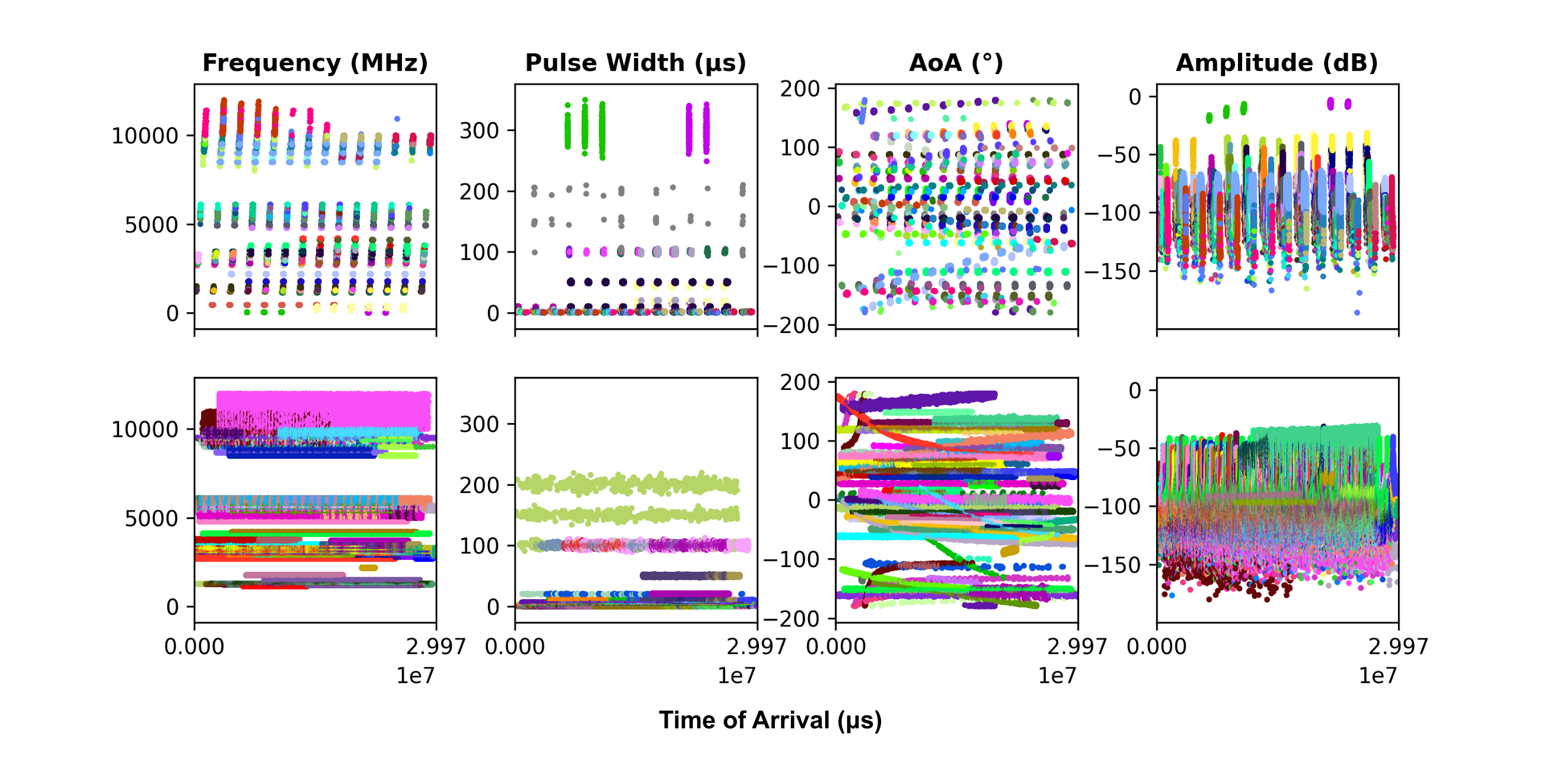}
    \caption[PDWs mimic realistic radar transmitters.]{\textbf{PDWs mimic realistic radar transmitters.} We simulated pulse transmission and detection in realistic environments characterised by 5-feature PDWs. Figure (A) and (B) demonstrate \textit{stare} and \textit{scan} receiver models over frequency, pulse width, AoA, and amplitude. The substantial overlap of radar pulses suggest that successful deinterleaving can only be achieved by leveraging temporal patterns over all parts of the PDWs.}
    \label{fig:Example-trains}
\end{figure*}

Simulated radio pulses were captured by a static, idealised receiver (Rx) and sent from transmitters (Tx) at varying, randomly sampled distances on a two-dimensional plane (Figure \ref{fig:summary-schematic}). To make the dataset mostly independent of Rx hardware characteristics, we focused on simulating realistic Tx properties and simplified signal detection. Data in the TSRD can be understood as the emitted ground truth in the environment rather than imitating receiver behaviour. To challenge model development, pulses were dropped when the Rx was not tuned to the correct frequency band, when the Tx was too far for detection, or when the pulse width dropped below a threshold (0.0069$\mu$s). Consequently, not all emitters were visible to the Rx. Pulse trains are provided in two Rx modes, one of which receives all signals over the entire possible frequency spectrum at any time over 30 seconds (\textit{stare} mode); the other is scanning through frequency bands in deterministic intervals (\textit{scan} mode). The \textit{stare} Rx model represents an oracle receiver that can observe the entire frequency spectrum at once, whereas the \textit{scan} mode  mimics real-world receivers. Both \textit{scan} and \textit{stare} produce highly interleaved pulse trains with associated emitter labels from up to 100 simulated emitters in train and validation set (Figure \ref{fig:trains-with-and-without-labels}). To challenge trained models during testing, we allowed up to 110 emitters when simulating the test set. Each receiver mode contains a total of 3000 pulse trains (train set $n=$ 2500; validation set $n=$ 250, hold out test set $n=$ 250) with varying complexity. The \textit{stare} Rx model captured up to 5,920,979 PDWs for training pulse trains (\textit{scan}: 505,094) with an average of 1,285,418.89 (\textit{scan}: 94,277.34) coming from up to 85 emitters (\textit{scan}: 90). All data statistics are provided in Table \ref{tab:dataset_statistics}, and we exemplify the pulse train data in Figure \ref{fig:Example-trains}. In the following sections, we explain the data generation in more detail.

\begin{table}[htbp]
    \centering
    \renewcommand{\arraystretch}{0.85}
    \setlength{\tabcolsep}{4pt}
\scriptsize
    \begin{tabular}{@{}llrrrr@{}}
        \toprule
        \textbf{Rx} & \textbf{Metric} & \textbf{Train} & \textbf{Val.} & \textbf{Test} & \textbf{All} \\
        \midrule
        \multirow{8}{*}{\textit{Stare}}
        & n trains       & 2,500   & 250     & 250     & 3,000  \\
        & Total pulses   & 3.17B   & 316.7M  & 367.5M  & 3.86B  \\
        & Max pulses     & 5.76M   & 5.92M   & 4.38M   & 5.92M  \\
        & Min pulses     & 0       & 91      & 1,587   & 0      \\
        & Mean pulses    & 1.27M   & 1.27M   & 1.47M   & 1.29M  \\
        & Max emitters   & 83      & 77      & 85      & 85     \\
        & Min emitters   & 0       & 1       & 1       & 0      \\
        & Mean emitters  & 36.7    & 36.0    & 43.3    & 37.2   \\
        \midrule
        \multirow{8}{*}{\textit{Scan}}
        & n trains       & 2,500   & 250     & 250     & 3,000  \\
        & Total pulses   & 233.2M  & 22.7M   & 27.0M   & 282.8M \\
        & Max pulses     & 390.5K  & 505.1K  & 354.8K  & 505.1K \\
        & Min pulses     & 0       & 4       & 103     & 0      \\
        & Mean pulses    & 93.3K   & 90.8K   & 107.9K  & 94.3K  \\
        & Max emitters   & 85      & 79      & 90      & 90     \\
        & Min emitters   & 0       & 1       & 1       & 0      \\
        & Mean emitters  & 38.1    & 37.1    & 44.3    & 38.5   \\
        \bottomrule
    \end{tabular}
    \vspace{4pt}
    \caption{Dataset statistics across train, validation, and test splits.}
    \label{tab:dataset_statistics}
\end{table}
\subsection{Dataset properties}
\begin{figure}
    \centering
    \includegraphics[width=0.85\linewidth]{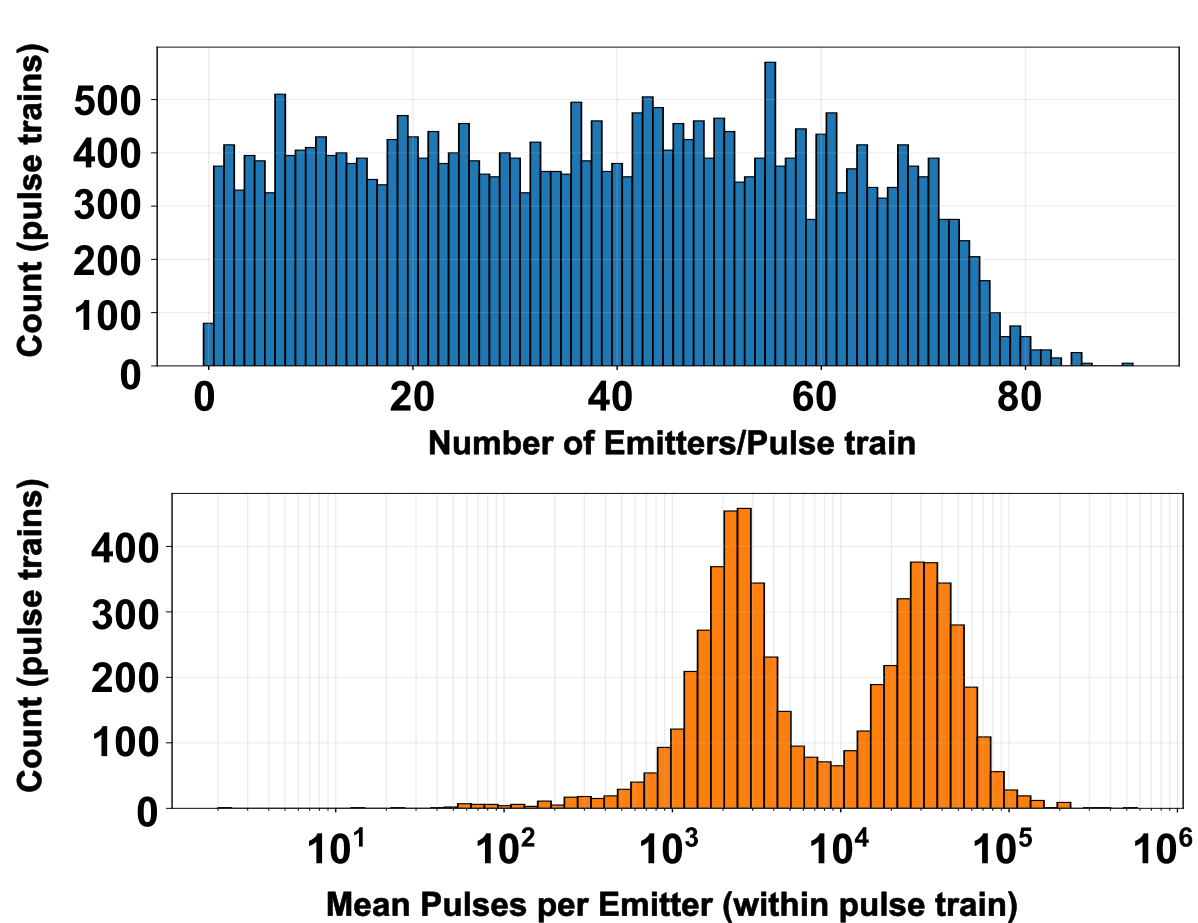}
    \caption[Emitter-level statistics are well balanced over the entire dataset.]{\textbf{Emitter-level statistics are well balanced over the entire dataset.} (Top) The number of emitters is approximately uniformly distributed over all pulse trains, rendering some more complex than others. Emitter numbers over 80 eventually tail off. (Bottom) As expected the average number of pulses per emitter follows a Poisson-like distribution as expected from count data. Statistics were computed in \textit{scan} mode.}
    \label{fig:emitter-stats}
\end{figure}

Pulse trains are PDW sequences of varying length with associated emitter labels. Each PDW is a five-dimensional vector composed of time of arrival (ToA), centre frequency, pulse width (PW), angle of arrival (AoA), and amplitude, reflecting typical radio transmission data. We simulated heavily congested environments by including pulses from up to 100 emitters per pulse train (110 for the test data). The TSRD contains both simple setups with low emitter numbers as well as complex scenarios with many sources, ensuring that algorithms need to perform on both. A detailed breakdown per PDW is provided in Tables \ref{tab:pdw_stats_stare} and \ref{tab:pdw_stats_scan} as well as in Figure \ref{fig:emitter-stats}.

\begin{table}[htp]
    \centering
    \renewcommand{\arraystretch}{0.85}
    \setlength{\tabcolsep}{4pt}
\scriptsize
    \begin{tabular}{@{}llrrrr@{}}
        \toprule
        \textbf{Split} & \textbf{PDW} & \textbf{Mean} & \textbf{Std} & \textbf{Min} & \textbf{Max} \\
        \midrule
        \multirow{5}{*}{Train}
            & ToA ($\mu$s)      & 14.34M & 7.16M & 175.61   & 43.49M \\
        &   Freq. (MHz)         & 4.63K      & 2.79K     & 998.62   & 16.09K  \\
        &   PW ($\mu$s)         & 5.77          & 19.27        & 0.007    & 232.78     \\
        &   AoA ($^\circ$)      & 10.33         & 96.81        & -180.00  & 180.00     \\
        &   Amp. (dB)           & -89.04        & 12.78        & -216.23  & 11.34      \\
        \midrule
        \multirow{5}{*}{Val.}
            & ToA ($\mu$s)      & 14.24M & 7.17M & 1.16K & 43.17M \\
        &   Freq. (MHz)         & 4.66K      & 2.88K     & 999.52   & 16.11K  \\
        &   PW ($\mu$s)         & 5.59          & 19.01        & 0.007    & 225.47     \\
        &   AoA ($^\circ$)      & 10.28         & 98.07        & -180.00  & 180.00     \\
        &   Amp. (dB)           & -88.93        & 12.87        & -212.55  & 10.97      \\
        \midrule
        \multirow{5}{*}{Test}
            & ToA ($\mu$s)      & 14.36M & 7.21M & 701.34   & 43.33M \\
        &   Freq. (MHz)         & 4.71K      & 2.81K     & 1.00K & 16.09K  \\
        &   PW ($\mu$s)         & 6.16          & 19.88        & 0.007    & 226.72     \\
        &   AoA ($^\circ$)      & 6.96          & 97.09        & -180.00  & 180.00     \\
        &   Amp. (dB)           & -89.06        & 12.95        & -207.67  & -4.69      \\
        \bottomrule
    \end{tabular}
    \vspace{4pt}
    \caption{PDW statistics per split in \textit{stare} mode.}
    \label{tab:pdw_stats_stare}
\end{table}
\begin{table}[htp]
    \centering
    \renewcommand{\arraystretch}{0.85}
    \setlength{\tabcolsep}{4pt}
\scriptsize
    \begin{tabular}{@{}llrrrr@{}}
        \toprule
        \textbf{Split} & \textbf{PDW} & \textbf{Mean} & \textbf{Std} & \textbf{Min} & \textbf{Max} \\
        \midrule
        \multirow{5}{*}{Train}
            & ToA ($\mu$s)      & 14.21M & 7.15M & 218.43   & 29.94M \\
        &   Freq. (MHz)         & 4.71K      & 3.13K     & 4.67     & 16.09K  \\
        &   PW ($\mu$s)         & 11.92         & 37.57        & 0.007    & 370.81     \\
        &   AoA ($^\circ$)      & 10.44         & 95.97        & -180.00  & 180.00     \\
        &   Amp. (dB)           & -86.12        & 17.17        & -212.499 & 34.14      \\
        \midrule
        \multirow{5}{*}{Val.}
            & ToA ($\mu$s)      & 14.08M & 7.18M & 1.49K & 30.14M \\
        &   Freq. (MHz)         & 4.90K      & 3.19K     & 4.79     & 16.07K  \\
        &   PW ($\mu$s)         & 11.62         & 37.28        & 0.007    & 368.25     \\
        &   AoA ($^\circ$)      & 12.57         & 96.85        & -180.00  & 180.00     \\
        &   Amp. (dB)           & -86.29        & 16.85        & -192.77  & 17.71      \\
        \midrule
        \multirow{5}{*}{Test}
            & ToA ($\mu$s)      & 14.21M & 7.18M & 3,396.46 & 29.53M \\
        &   Freq. (MHz)         & 4.88K      & 3.15K     & 4.68     & 16.07K  \\
        &   PW ($\mu$s)         & 11.33         & 36.39        & 0.007    & 373.27     \\
        &   AoA ($^\circ$)      & 6.53          & 95.12        & -180.00  & 180.00     \\
        &   Amp. (dB)           & -86.46        & 17.03        & -193.26  & 23.99      \\
        \bottomrule
    \end{tabular}
    \vspace{4pt}
    \caption{PDW statistics per split in \textit{scan} mode.}
    \label{tab:pdw_stats_scan}
\end{table}
\par Sent signals imitate physical properties of in total 68 transmitter \textit{types}, which define hardware constraints, realistic parameter ranges and modulation schemes.
%including the \textit{Furuno FAR-2238S X-Band Marine Radar} or \textit{ASDE-X Airport Surface Detection Radar}.
Individual transmitter \textit{instances} were then randomly sampled and initiated with different operating modes, e.g. static configuration, frequency hopping, staggered PRI, or other advanced techniques. Tx instances could move along straight lines at arbitrary but constant velocities in a 2-dimensional plane. Initial positions were uniformly sampled within a 250-by-250 km range. 

\par Transmission and measurement uncertainty was incorporated via specialised noise models that build on a line-of-sight path loss without multi-path interference, representing idealised but realistic setups. In particular, we combined additive Gaussian White Noise (GWN) with an adapted mean-reverting Ornstein-Uhlenbeck (OU) noise process for modelling signal decay and interference depending on the modelled signal property. During pulse emission, ToA and pulse width were adjusted with additive GWN, whereas frequencies were jittered using the OU noise model. The receiver independently added GWN to ToA and pulse width, and amplitude and AoA were blurred using the OU process to model atmospheric interference. The received frequency was not adjusted. Collectively, we considered idealised but authentic physical and environmental noise factors to bridge, as much as possible, the simulation-to-real data knowledge gap.  

\begin{figure*}[ht]
    \centering
    \includegraphics[width=0.8\linewidth]{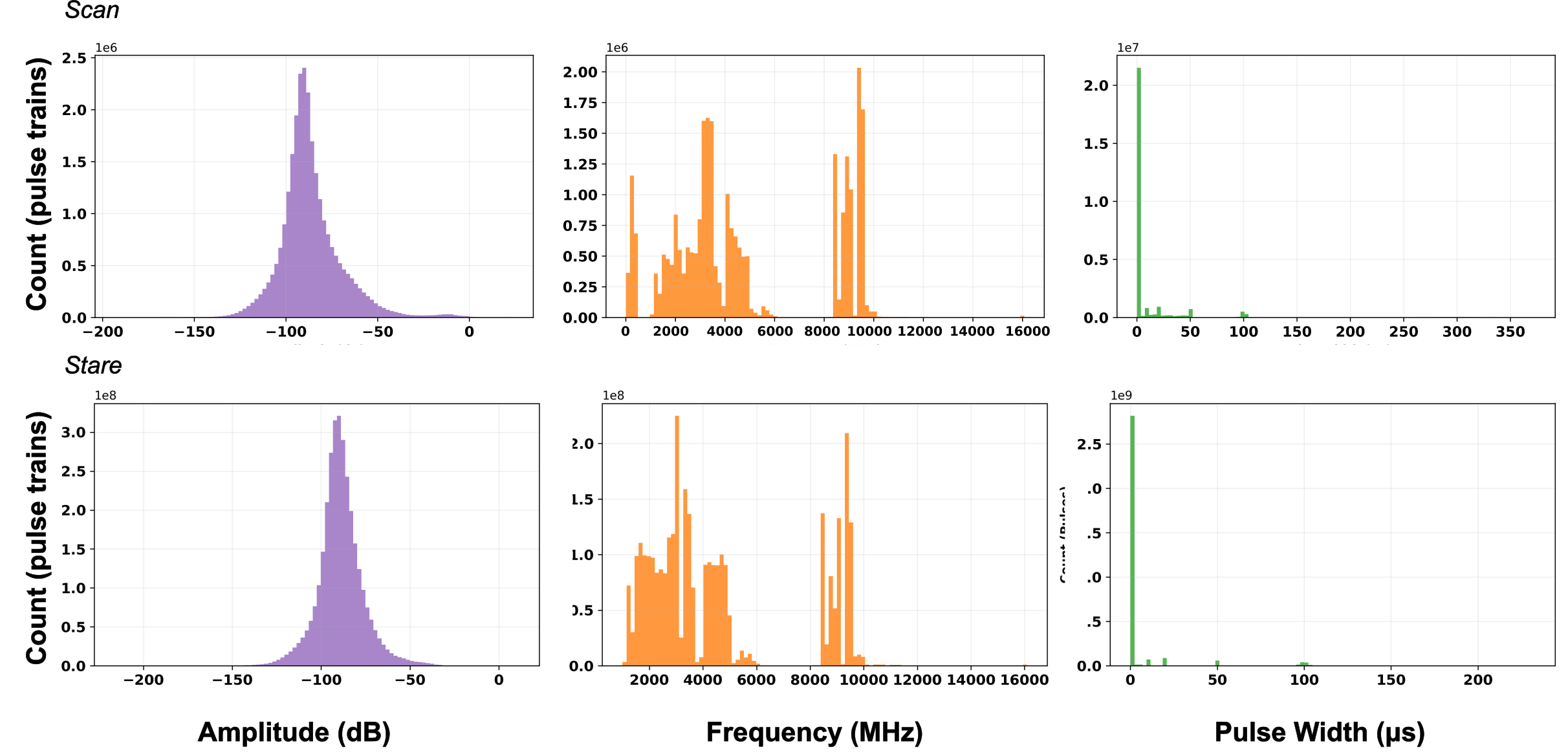}
    \caption[Distributions for amplitude, frequency, and pulse width.]{\textbf{Distributions for amplitude, frequency, and pulse width.} PDWs are differently distributed across pulse trains, as demonstrated for amplitude, frequency, and pulse width (left to right) for \textit{scan} mode (top) and \textit{stare} model (bottom).}
    \label{fig:parameter-dists}
\end{figure*}

\par We simulated two receiver models at fixed but arbitrary positions, which we refer to as \textit{scan} and \textit{stare} mode. In \textit{stare} mode, the receiver detects all pulses with frequencies of up to 18 GHz. Collection time was set to 30$s$. The simulation has an ambient noise of -100 dB. The received amplitude decreases quadratically with emitter distance, and the probability of pulse detection increases the more distinct the signal is from the noise floor. The receiver was oriented in a static angle with an antenna gain of 10 dB. As emitters can transmit simultaneously on different frequencies, pulses can have identical ToAs, for which conventional pulse-by-pulse deinterleavers might fail. Aiming to provide a realistic  scenario, we implemented the \textit{scan} receiver model which sweeps the frequency spectrum at centre frequencies between 0.5 - 18 GHz in 500 MHz steps and 500 MHz bandwidth at deterministic but varying dwell times. Pulses sent on frequencies outside the tuned 500 MHz bandwidth were dropped, which is why the scan mode only shows pulses in a given frequency range in a given interval. All other receiver parameters remain unchanged.

% \par To ensure robustness, pulses are randomly dropped as a function of signal-to-noise ratio. Signals can only be detected by the receiver if configured to the correct frequency, which is why the \textit{scan} mode only shows pulses in a given frequency range, in an interval (see Figure \ref{fig:parameter-dists}).

\par We introduced a substantial label imbalance with some pulse trains being represented by the same emitter at a proportion of up to 99.7\%, which leads to a median per-emitter contribution of 2.4\%. Whilst this is intended to mimic realistic scenarios which most current deinterleaving models struggle with, the high proportion of strongly dominating transmitters is likely exaggerated.

\par
\begin{figure}[ht]
    \centering
    \includegraphics[width=0.7\linewidth]{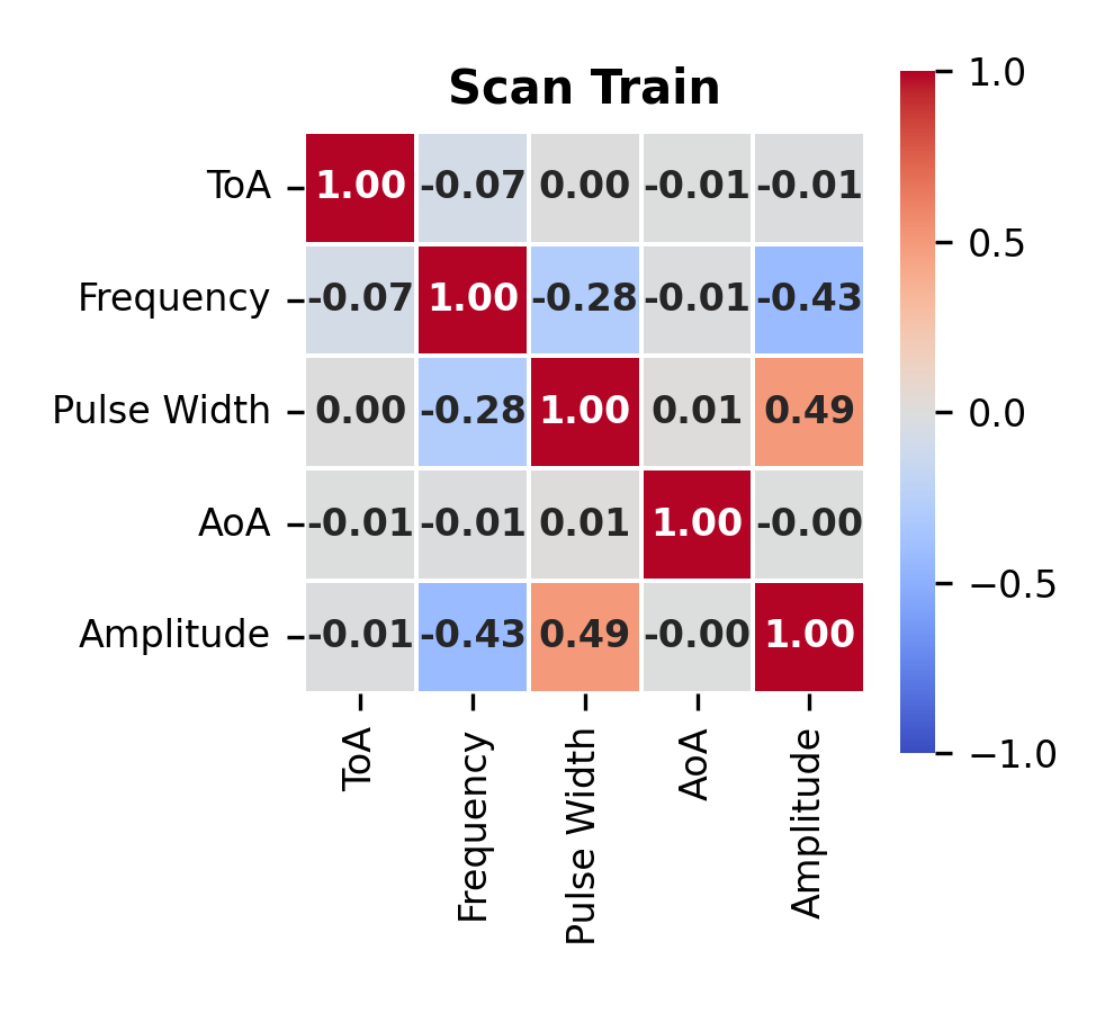}
    \caption[PDW features are largely independent, suggesting that every feature can contribute to better task performance.]{\textbf{PDW features are largely independent, suggesting that every feature can contribute to better task performance.} Although frequency exhibits a weak correlation with pulse width and amplitude, most PDW features are independent of each other, indicating that all data properties can contribute useful statistics for downstream tasks. Correlations were measured for \textit{scan} mode.}
    \label{fig:correlations}
\end{figure}

 Analysing distributions (Figure \ref{fig:parameter-dists}) and Spearman correlations (Figure \ref{fig:correlations}) between PDWs suggests that features are mostly independent of each other, although some moderate positive and negative correlations exist due to physical constraints (i.e., amplitude, frequency, and pulse width, Figure \ref{fig:correlations}). No subset of these properties is sufficient for explaining any other PDW characteristic. These data therefore require the use of methods which extract higher-order patterns to deinterleave.

\par To summarise, the TSRD is the first publicly available, synthetically generated, comprehensive dataset with varying complexity, mimicking ideal and real-wold scenarios for model development on pulse trains represented as PDWs.

\subsection{The Turing Deinterleaving Challenge}
\begin{table}[htp]
    \centering
    \begin{tabular}{@{}l*{7}{r}@{}}
        \toprule
        \textbf{Rx} & \textbf{V} & \textbf{ARI} & \textbf{AMI} & \textbf{H} & \textbf{C} & \textbf{MCC} & \textbf{F1} \\
        \midrule
        \textit{Stare} & 0.54 & 0.27 & 0.50 & 0.64 & 0.50 & 0.06 & 0.01 \\
        \textit{Scan} & 0.19 & 0.02 & 0.15 & 0.41 & 0.13 & 0.07 & 0.04 \\
        \bottomrule 
        \newline
    \end{tabular}
    \caption{HDBscan clustering performance on the raw PDWs of the test data. \textbf{V}: V-measure, \textbf{ARI}: Adjusted Rand Index, \textbf{AMI}: Adjusted Mutual Information, \textbf{H}: Homogeneity, \textbf{C}: Completeness, \textbf{MCC}: Matthew's Correlation Coefficient.}
    \label{tab:baseline}
\end{table}

We propose the \textit{Turing Deinterleaving Challenge}, which aims to maximise median cluster metrics, particularly V-measure, Adjusted Rand Index (ARI), Adjusted Mutual Information (AMI), homogeneity, and completeness, across the test dataset. The number of emitters per window is unknown at test time and varies between windows, reflecting realistic operational conditions. Participants must handle significant overlap in the parameter space between different emitters, realistic noise conditions, and varying pulse train characteristics. The challenge includes emitters with different behaviours, from simple constant-PRI radars to more complex agile systems with frequency hopping and variable pulse repetition intervals. As it is known that clustering metrics can over-estimate performance on strongly skewed label distributions, we also include pairwise-binary metrics, such as Matthew's Correlation Coefficient (MCC) and F1 score. Pairwise-binary metrics are determined by calculating their score based on all true and predicted label pairs, followed by taking the maximum over the true labels (for identifying corresponding labels) and the minimum over the predicted labels. Consequently, the pairwise-binary metrics can be seen as an average worst-case performance. We provide a Python library for loading, windowing, and saving the data as well as running the benchmark metrics against models to be tested (\url{https://github.com/alan-turing-institute/turing-deinterleaving-challenge}). The library aims to facilitate interfacing with modern machine learning pipelines, such as \cite{gunn_radar_2025}.

\par
\begin{figure}[ht]
    \centering
    \includegraphics[width=0.8\linewidth]{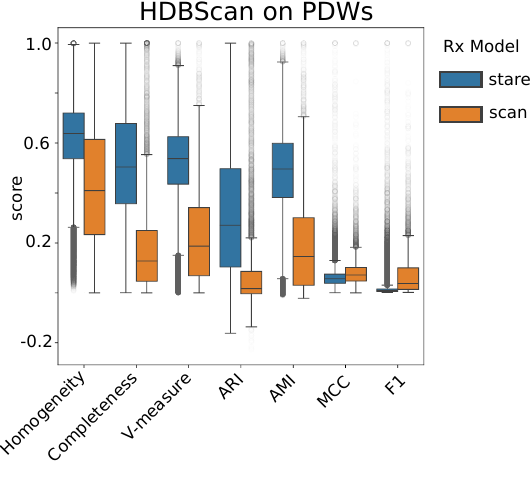}
    \caption[HDBscan on the raw PDWs provides a first baseline for the \textit{Turing Deinterleaving Challenge}.]{\textbf{HDBscan on the raw PDWs provides a first baseline for the \textit{Turing Deinterleaving Challenge}.} Whilst HDBscan cluster of pulses collected in \textit{stare} yields higher values for V-measure, AMI, ARI, homogeneity, and completeness, \textit{scan} yields a slightly better worst-case performance as measured in the pairwise-bianry metrics MCC and F1.}
    \label{fig:baseline}
\end{figure}

 To provide a baseline, we applied HDBScan clustering \cite{HDBSCAN} directly on the non-transformed PDWs of the test \textit{stare} and \textit{scan} dataset on non-overlapping 1024 pulse-sized windows. The algorithm hierarchically identifies data point densities in the data and merges found agglomerations if their distance is less than a threshold $\epsilon$. After evaluating $\epsilon \in \{0, 0.05, 0.1, 0.5\}$ on a reduced dataset of a 1000 windows, we discovered that all parameters performed equivalently for both receiver models, and we arbitrarily selected $\epsilon = 0$. Whilst HDBscan clustering yields a V-measure of 0.54 on the \textit{stare} dataset, it performs poorly when clustering pulses in \textit{scan} mode, with a V-measure of only 0.19 (Figure \ref{fig:baseline} and Table \ref{tab:baseline}). On the other hand, HDBscan achieves a marginally better pairwise-binary MCC (0.071) and F1 score (0.037) on \textit{scan} than on \textit{stare} Rx data (0.057 and 0.010 MCC and F1 on \textit{stare}, respectively). Overall, this demonstrates that more sophisticated models are needed for successful deinterleaving.

\section{Outlook \& Conclusion}
\label{sec:outlook}
The TSRD provides the radar deinterleaving research community, which has historically relied on proprietary data, with its first public, large-scale, and model-agnostic benchmark dataset. We designed the TSDR to address key limitations in existing deinterleaving research, including the assumption of fixed emitter numbers, limited use of available PDW features, and lack of standard evaluation procedures. The accompanying \textit{Turing Deinterleaving Challenge} provides a simple interface for measuring model performance. The TSDR represents a first step towards an accelerated community-driven model development effort and the establishment of public best-practice benchmarks for EW research.
\bibliographystyle{IEEEtran}
\bibliography{references}

\end{document}